\documentstyle[preprint,aps]{revtex}
\newcommand{\sumv}{\sum_v}
\newcommand{\tr}[1]{\mbox{\rm tr} \left\{ #1 \right\}}
\newcommand{\Lm}{\gamma_\mu (1-\gamma_5)}
\begin{document}
\draft
\preprint{DPNU-93-37, OS-GE 38-93}
\title{Heavy meson effective theory with $1/M_Q$ correction}
\author{Noriaki Kitazawa
 \thanks{JSPS fellow. {\bf e-mail}:
         {\tt kitazawa@eken.phys.nagoya-u.ac.jp}}}
\address{Department of Physics, Nagoya University,\\
         Nagoya 464-01, Japan}
\author{Takeshi Kurimoto
 \thanks{{\bf e-mail}: {\tt krmt@fuji.wani.osaka-u.ac.jp}}}
\address{Institute of Physics, College of General Education,
         Osaka University,\\
         Toyonaka 560, Osaka, Japan}
\date{\today}
\maketitle
\begin{abstract}
We construct
 an effective Lagrangian of heavy and light mesons
 with $1/M_Q$ correction.
The Lagrangian is constructed model independent way
 by using only the information of the symmetry of QCD.
Reparameterisation invariance at the meson level
 is taken into account for the consistency of the theory.
The partial decay width of the process
 $D^{*+} \rightarrow D^0 \pi^+$
 and the form factors of the process
 ${\bar B}^0 \rightarrow \pi^+ l {\bar \nu}$
 are calculated with the $1/M_Q$ correction.
We also introduce the light vector mesons
 based on the approach of the hidden local symmetry.
The form factors of the process
 ${\bar B}^0 \rightarrow \rho^+ l {\bar \nu}$ are calculated
 with thus introduced $\rho$-meson.
These results are easily translated
 to the $D$-meson semileptonic decays.
\end{abstract}
\newpage

The spin-flavor symmetry in the heavy quark sector\cite{Isgur-Wise}
 has been extensively studied to understand the heavy hadron decays.
Since the proposed projects
 ($B$-factories, $\tau$-charm factories and so on)
 are mainly accessible for the heavy mesons,
 we concentrate on the physics of the heavy mesons in this letter.
There are two ways to extract the consequences of the symmetry.
One is based on the heavy quark effective fields
 (heavy quark effective theory)\cite{Heavy-quark},
 and another is based on the heavy meson effective fields
 (heavy meson effective theory)\cite{Heavy-meson}.
The algebra of the currents and charges
 in the heavy quark effective theory
 is used to reduce the number of independent form factors
 in the weak current matrix elements.
It is well known that the six form factors
 in $B \rightarrow D$ and $B \rightarrow D^*$
 weak transition matrix element
 are described by a single function called Isgur-Wise function
 in the heavy mass limit\cite{Isgur-Wise}.
The $1/M_Q$ correction and the QCD correction
 of the heavy-to-heavy weak current matrix element
 have been calculated\cite{Luke}.
The reparameterisation invariance\cite{Luke-Manohar}
 plays an important role in the calculation.
The result contains some form factors as unknown functions.

The construction of the heavy meson effective Lagrangian
 is based on both chiral symmetry
 in the light quark (u,d, and s quarks) sector
 and the spin-flavor symmetry
 in the heavy quark (c and b quarks) sector\cite{Heavy-meson}.
The semileptonic decays of the heavy mesons have been treated
 in leading order of the $1/M_Q$ expansion
 by using the heavy meson effective Lagrangian
 \cite{Heavy-meson,Casalbuoni-et-al-1,Yan-et-al}.
It should be noted that
 there is a $B^*$ pole contribution in the form factors
 of $B \rightarrow \pi l \nu$,
 which is similar to the $\rho$ pole contribution
 in the form factor of $\pi$.
The $B^*$ meson is naturally introduced
 as the spin partner of the $B$ meson.
Light vector mesons are introduced
 in the heavy meson effective Lagrangian\cite{Casalbuoni-et-al-2}
 based on the hidden local symmetry\cite{Bando-et-al}.

In this paper
 we construct the heavy meson effective Lagrangian
 up to the $O(1/M_Q^2)$ corrections in $1/M_Q$ expansion
 and $O(p^2)$ in the chiral expansion
 \footnote{\scriptsize
           H-Y.Cheng et al.
            have also considered the $1/M_Q$ correction
            in heavy meson effective Lagrangian\cite{Cheng-et-al}.
           But they have not emunerated
            all possible $O(1/M_Q$) terms
            in terms of the effective meson field $H_v$,
            and also have not considered the $1/M_Q$ correction
            in weak currents.}.
The explicit chiral symmetry breaking is worth studying,
 but it is left for our future works.
Since the construction is based only on the symmetry alone,
 the results from the Lagrangian should be model independent.
The Lagrangian contains ten parameters
 (six couplings, two heavy quark masses,
 a mass scale of the order of $\Lambda_{QCD}$
 (typical scale of the QCD dynamics),
 and a parameter which describe the mass splitting
 between the heavy pseudoscalar and vector mesons)
 which should be fixed by the experiments.
The decay width of the process $D^{*+} \rightarrow D^0 \pi^+$
 and the form factors of the process
 ${\bar B}^0 \rightarrow \pi^+ l {\bar \nu}$
 are calculated with the $1/M_Q$ correction.
The light vector mesons are introduced
 according to the method of the hidden local symmetry.
The form factors of the process
 ${\bar B}^0 \rightarrow \rho^+ l {\bar \nu}$ are also calculated
 with thus introduced light vector mesons.

The Lagrangian is constructed by the following effective fields.
The heavy mesons are described by the field
\begin{equation}
 H_v = {{1 + \not\!v} \over 2}
       \left[ i \gamma_5 P_v + \gamma_\mu P^{*\mu}_v \right],
\end{equation}
 where $v$ is the velocity of the heavy quark inside.
The meson momentum $p$ is described by
\begin{equation}
 p = M_Q v + k,
\end{equation}
 where $k$ is the residual momentum of the order of $\Lambda_{QCD}$.
The fields $P_v$ and $P^*_v$, which have mass dimension $3/2$,
 are the heavy pseudoscalar and heavy vector fields, respectively.
They are the multiplets in quark flavor given as
\begin{equation}
 P_v^{(*)} =
 \left(
  \begin{array}{ccc}
   D^0 & D^+        & D^+_s        \\
   B^- & {\bar B}^0 & {\bar B}^0_s
  \end{array}
 \right)^{(*)}.
\end{equation}
The field $H_v$ is transformed
 under the spin-flavor $SU(4)$ transformation
 which is decomposed by $SU(2)_{spin}$ and $SU(2)_H$ transformation
 and chiral transformation ($SU(3)_L \times SU(3)_R$ transformation)
 as
\begin{eqnarray}
 H_v &\rightarrow& S H_v,
\\
 H_v &\rightarrow& z_H H_v,
\\
 H_v &\rightarrow& H_v h(\Pi, g_L, g_R)^{\dag},
\end{eqnarray}
 where $S \in SU(2)_{spin}$ acts on the Dirac index,
 $z_H \in SU(2)_H$ on the heavy flavor index.
The chiral transformation is non-linearly realized as
\begin{equation}
 \xi \rightarrow g_L \xi h(\Pi, g_L, g_R)^{\dag}
               = h(\Pi, g_L, g_R) \xi g_R,
\end{equation}
 where $g_L \in SU(3)_L$, $g_R \in SU(3)_R$,
 and $\xi$ is defined in terms of the field
 of the Nambu-Goldstone bosons
 ($\pi$, $K$, $\eta_8$),
\begin{equation}
 \xi = e^{i\Pi/f_{\pi}}
\qquad\qquad
 \left( \Pi = \Pi^a {{\lambda^a} \over 2} \right).
\end{equation}
The unitary matrix $h(\Pi,g_L,g_R)$ has a complicated form,
 but if we consider the vector transformation $g=g_L=g_R$,
 then $h(\Pi,g_L,g_R)=g$.
For the convenience to consider the chiral expansion
 (derivative expansion),
 the followings are used to build the Lagrangian,
\begin{eqnarray}
 \alpha_\perp^\mu
  &=& {i \over 2} \left( \xi \partial^\mu \xi^{\dag}
                       - \xi^{\dag} \partial^\mu \xi \right),
\\
 \alpha_\parallel^\mu
  &=& {i \over 2} \left( \xi \partial^\mu \xi^{\dag}
                       + \xi^{\dag} \partial^\mu \xi \right).
\end{eqnarray}
These are transformed under the chiral transformation as
\begin{eqnarray}
 \alpha_\perp^\mu &\rightarrow& h \alpha_\perp^\mu h^{\dag},
\\
 \alpha_\parallel^\mu &\rightarrow&
  h \alpha_\parallel^\mu h^{\dag} + h i \partial^\mu h^{\dag}.
\end{eqnarray}

The parity transformation and charge conjugation
 are defined as follows.
The effective fields are transformed
 under the parity transformation as
\begin{eqnarray}
 {\cal P} H_v(x) {\cal P}^{\dag}
  &=& \gamma^0 H_{\bar v}({\bar x}) \gamma^0,
\\
 {\cal P} \alpha_\perp^\mu(x) {\cal P}^{\dag}
  &=& - \alpha_{\perp\mu}({\bar x}),
\\
 {\cal P} \alpha_\parallel^\mu(x) {\cal P}^{\dag}
  &=& \alpha_{\parallel\mu}({\bar x}),
\end{eqnarray}
 where ${\bar x} = (x^0, - \mbox{\boldmath $x$})$
 and ${\bar v} = (v^0, - \mbox{\boldmath $v$})$,
 and under the charge conjugation
\begin{eqnarray}
 {\cal C} H_v(x) {\cal C}^{\dag}
  &=& C \left( \overline{H_v^{(-)}}(x) \right)^TC^{\dag},
\\
 {\cal C} \alpha_\perp^\mu(x) {\cal C}^{\dag}
  &=& \left( \alpha_\perp^\mu(x) \right)^T,
\\
 {\cal C} \alpha_\parallel^\mu(x) {\cal C}^{\dag}
  &=& - \left( \alpha_\parallel^\mu(x) \right)^T,
\end{eqnarray}
 where ${\cal P}$ and ${\cal C}$ are the operators,
 $C = i \gamma^2 \gamma^0$,
 and $H_v^{(-)}$ is the effective field of the heavy mesons
 which contain the negative energy component of the heavy quark.
The field $H_v^{(-)}$ is transformed in the same way as $H_v$
 under the spin and flavor transformation.

Now the effective Lagrangian can be obtained
 by imposing the chiral symmetry
 and the invariance under the parity transformation
 and charge conjugation.
Spin-flavor symmetry is imposed
 with the breaking terms of $O(1/M_Q)$ included.
We generally write down all the possible terms and get
\begin{eqnarray}
 {\cal L} &=&
 - \sumv \tr{ {\bar H}_v v \cdot iD H_v }
 - \sumv \tr{ {\bar H}_v {{(iD)^2} \over {2M}} H_v }
\nonumber\\
&&
 + \Lambda \sumv \tr{ {\bar H}_v H_v }
 + \kappa' \Lambda \sumv \tr{ {\bar H}_v {\Lambda \over M} H_v }
 + \kappa \Lambda \sumv
   \tr{ {\bar H}_v {\Lambda \over M} \sigma_{\rho\sigma}
                                 H_v \sigma^{\rho\sigma}}
\nonumber\\
&&
 + r \sumv \tr{ {\bar H}_v H_v v \cdot {\hat \alpha}_\parallel }
\nonumber\\
&& \qquad
 + r \sumv \tr{ {\bar H}_v {{iD_\mu} \over {2M}}
                H_v \alpha_\parallel^\mu } + h.c.
\nonumber\\
&& \qquad
 + r_1 \sumv
  \tr{ {\bar H}_v {\Lambda \over M} H_v
       v \cdot {\hat \alpha}_{\parallel} }
 + r_2 \sumv
  \tr{ {\bar H}_v {\Lambda \over M} \sigma^{\rho\sigma} H_v
       \sigma_{\rho\sigma} v \cdot {\hat \alpha}_{\parallel} }
\nonumber\\
&&
 + \lambda \sumv \tr{ {\bar H}_v H_v
                      \gamma_\mu \gamma_5 \alpha_\perp^\mu }
\nonumber\\
&& \qquad
 - \lambda \sumv \tr{ {\bar H}_v {{v \cdot iD} \over {2M}} H_v
                      \gamma_\mu \gamma_5 \alpha_\perp^\mu } + h.c.
\nonumber\\
&& \qquad
 - \lambda \sumv \varepsilon^{\mu\nu\rho\sigma}
  \tr{ {\bar H}_v {{iD_\rho} \over {4M}} H_v
       \sigma_{\mu\nu} \alpha_{\perp\sigma} }
 + \mbox{h.c.}
\nonumber\\
&& \qquad
 + \lambda_1 \sumv
  \tr{ {\bar H}_v {\Lambda \over M} H_v
       \gamma_\rho \gamma_5 \alpha_\perp^\rho }
 + \lambda_2 \sumv
  \tr{ {\bar H}_v {\Lambda \over M} \gamma_\rho \gamma_5 H_v
       \alpha_\perp^\rho }
\nonumber\\
&&
 + \mbox{(Anti-particle)},
\end{eqnarray}
 where $1/M=\mbox{diag}(1/M_c, 1/M_b)$.
Anti-particle part has exactly the same form of the particle part but
 $H_v \rightarrow H_v^{(-)}$ and $v \rightarrow -v$.
The pseudoscalar meson masses $m_P$ and the vector meson masses $m_V$
 are expanded as
\begin{eqnarray}
 m_P^2 &=& M_Q^2 \left\{ 1 + 2 {\Lambda \over {M_Q}}
                           + 2 \kappa' {{\Lambda^2} \over {M_Q^2}}
                           + 12 \kappa {{\Lambda^2} \over {M_Q^2}}
                 \right\},
\nonumber\\
 m_V^2 &=& M_Q^2 \left\{ 1 + 2 {\Lambda \over {M_Q}}
                           + 2 \kappa' {{\Lambda^2} \over {M_Q^2}}
                           - 4 \kappa {{\Lambda^2} \over {M_Q^2}}
                 \right\}.
\end{eqnarray}
It should be noted that
 the normalisation of the effective field is different from
 that of the conventional boson fields by $\sqrt{M_Q}$.
Light vector mesons are introduced
 by the method of the hidden local symmetry\cite{Bando-et-al}.
The covariant derivative in the Lagrangian is defined as
\begin{equation}
 iD_\mu H_v = i\partial_\mu H_v - H_v g_V V_\mu,
\end{equation}
 where $V_\mu = V^a_\mu \lambda^a/2$ is the light vector meson field,
 $g_V$ is the gauge coupling constant of the hidden local symmetry,
 and ${\hat \alpha}_\parallel^\mu
      \equiv \alpha_\parallel^\mu - g_V V^\mu$.
We used the equation of motion
 to drop out the several terms
 as the higher order terms in $1/M_Q$ expansion and chiral expansion.
We take the reparameterisation invariance into account
 up to $O(1/M_Q^2)$ at the meson level.
The reparameterisation transformation of the effective field is
\begin{eqnarray}
 H_v &\rightarrow&
  \Lambda(v,{{w+iD/M} \over {|w+iD/M|}})
  \Lambda({{w+iD/M} \over {|w+iD/M|}},w)
  H_w
  \Lambda(w,{{w+i{\overleftarrow D}/M}
       \over {|w+i{\overleftarrow D}/M|}})
  \Lambda({{w+i{\overleftarrow D}/M}
       \over {|w+i{\overleftarrow D}/M|}},v)
  e^{-iqx}
\nonumber\\
 &=& \left\{ H_w - {1 \over {2M}} [ \not\!q, H_w ] \right\} e^{-iqx}
     +O(1/M^2),
\end{eqnarray}
 where $w=v+q/M$ and
\begin{equation}
 \Lambda(w,v) = {{1 + \not\!w \not\!v} \over \sqrt{2(1+w \cdot v)}}.
\end{equation}
Other fields are not transformed.

Many terms in the effective Lagrangian
 are dropped out by the reparameterisation invariance as follows:
For example, we can consider the term
\begin{equation}
 \tr{ {\bar H}_v {{iD^\rho} \over {M}} H_v
      \gamma_\rho \gamma_5 v \cdot \alpha_\perp }.
\end{equation}
The reparameterisation invariant form of this term is
\begin{equation}
 \tr{ {\bar {\tilde H}}_v {\cal V}^\rho {\cal V}^\sigma {\tilde H}_v
      \gamma_\rho \gamma_5 \alpha_{\perp\sigma} },
\label{exam-1-repara}
\end{equation}
 where ${\cal V} = v + iD/M$ and
\begin{eqnarray}
 {\tilde H}_v &=&
  \Lambda({{v+iD/M} \over {|v+iD/M|}},v) H_v
  \Lambda(v,{{v+i{\overleftarrow D}/M}
      \over {|v+i{\overleftarrow D}/M|}})
\nonumber\\
 &=& H_v + {1 \over {2M}}
     \left\{ \left[ \gamma_\mu, iD^\mu H_v \right]
             - 2 v \cdot iD H_v \right\} + O(1/M^2).
\end{eqnarray}
Since ${\tilde H}_v$ satisfies the relations
\begin{eqnarray}
 \not\!{\cal V} {\tilde H}_v &=& {\tilde H}_v,
\\
 {\tilde H}_v \not\!{\cal {\overleftarrow V}} &=& - {\tilde H}_v,
\end{eqnarray}
 the term (\ref{exam-1-repara}) vanishes.
Another example is the term
\begin{equation}
 \epsilon^{\alpha\beta\rho\sigma}
 \tr{ {\bar H}_v {{iD_\alpha} \over M} \sigma_{\rho\sigma} H_v
      \alpha_{\perp\beta} }.
\label{exam-2}
\end{equation}
The reparameterisation invariant form of this term is
\begin{equation}
 \epsilon^{\alpha\beta\rho\sigma}
 \tr{ {\bar {\tilde H}}_v {\cal V}_\alpha
       \sigma_{\rho\sigma} {\tilde H}_v \alpha_{\perp\beta} }.
\end{equation}
The leading component
\begin{equation}
 \epsilon^{\alpha\beta\rho\sigma}
 \tr{ {\bar H}_v v_\alpha
      \sigma_{\rho\sigma} H_v \alpha_{\perp\beta} }
\end{equation}
 which should be of the order of $1/M_Q$,
 because it breaks the spin symmetry.
So the original term (\ref{exam-2}) is $O(1/M_Q^2)$.
(The leading $O(1/M_Q)$ component reduces to the $\lambda_2$ term.)

We can do in the same way for the weak currents.
The heavy-to-light weak current is obtained
 up to the $O(1/M_Q^2)$ corrections in $1/M_Q$ expansion
 and $O(p^2)$ in chiral expansion as
\begin{eqnarray}
 J_\mu^{ia}(0) &=& F \biggl[ \tr{ (\xi^{\dag})^{ji} \Lm H_v^{aj} }
\nonumber\\
&& \qquad\qquad
   + {1 \over {2M_a}}
     \tr{ (\xi^{\dag})^{ji} \Lm [\gamma_\rho, iD^\rho H_v^{aj}] }
  \biggr]
\nonumber\\
&&
 + \alpha_1 {\Lambda \over {M_a}}
   \tr{ (\xi^{\dag})^{ji} \Lm H_v^{aj} }
\nonumber\\
&&
 + \alpha_2 {\Lambda \over {M_a}}
   \tr{ (\xi^{\dag})^{ji} \Lm \gamma^\rho H_v^{aj} \gamma_\rho }
\nonumber\\
&&
 + \beta_1 \Lambda^{1/2}
   \tr{ (\xi^{\dag})^{ji} \Lm H_v^{aj}
        \left( v \cdot {\hat \alpha}_\parallel \right) }
\nonumber\\
&&
 + \beta_2 \Lambda^{1/2}
   \tr{ (\xi^{\dag})^{ji} \Lm H_v^{aj}
        \left( \gamma_\rho {\hat \alpha}_\parallel^\rho \right) }.
\end{eqnarray}
This current contains $L_\mu = \gamma_\mu (1-\gamma_5)$.
We regard it in the construction of the current
 as an external field which is transformed
 under the spin-flavor transformation as
\begin{eqnarray}
 L_\mu &\rightarrow& g_L L_\mu z_H^{\dag},
\\
 L_\mu &\rightarrow& L_\mu S^{\dag}.
\end{eqnarray}
We also impose the ``parity'' invariance under the transformation
\begin{equation}
 L_\mu \rightarrow \gamma^0 L_\mu \gamma^0
\end{equation}
 to keep the $V-A$ structure of the current.

{}From this current,
 we can extract the decay constants of the heavy mesons.
\begin{eqnarray}
 f_P &=& \sqrt{2 \over {M_Q}}
         \left\{ F + {\Lambda \over {M_Q}}
                ( \alpha_1 + 2 \alpha_2) \right\},
\\
 f_V &=& \sqrt{2 \over {M_Q}}
         \left\{ F + {\Lambda \over {M_Q}}
                ( \alpha_1 - 2 \alpha_2) \right\}.
\end{eqnarray}
The heavy-to-heavy current can also be obtained in the same way,
 and the result contains some functions of $v \cdot v'$
 instead of the parameters,
 where $v$ and $v'$ are the velocities of the heavy quarks.
Weak interaction
 is introduced as the contact interaction between the currents
 \footnote{Here, the QCD correction is not considered
            for the simplicity in the presentation
            of the Lagrangian and current construction.}.

We get the width of the $D^{*+} \rightarrow D^0 \pi^+$ decay
 from the Lagrangian.
Since the energy-momentum of $\pi$ is very small
 (in the $D^*$ rest frame),
 our Lagrangian is accessible for this process.
We can use the axial vertex in the Lagrangian
 (couplings $\lambda$, $\lambda_1$, and $\lambda_2$).
The width is calculated as
\begin{equation}
 \Gamma(D^{*+} \rightarrow D^0 \pi^+)
 = {{\lambda^2 M_c^2 \left( E_\pi^2 - m_\pi^2 \right)^{3/2}}
                          \over {12\pi m_{D^*}^2 f_\pi^2}}
   \left[ 1 + {{E_\pi} \over {M_c}}
        + {{2 \left( \lambda_1 - \lambda_2 \right)} \over \lambda}
          {\Lambda \over {M_c}}
   \right],
\end{equation}
 where
\begin{equation}
 E_\pi = {{m_{D^*}^2 - m_D^2 + m_\pi^2} \over {2 m_{D^*}}}.
\end{equation}
The term $E_\pi / M_c$ can be neglected because it is $O(1/M_c^2)$
 due to the kinematics ($E_\pi = O(\Lambda^2/M_c)$).
But it should be kept in the off-shell amplitude
 as we will see next in $B \rightarrow \pi l \nu$ decay.
The couplings $\lambda_1$ and $\lambda_2$
 are contained in the combination of $\lambda_1-\lambda_2$.
The branching ratio of the decay
 is already measured by CLEO\cite{CLEO}.
If the total width is fixed,
 we can constraint the parameters
 $\lambda$ and $\lambda_1-\lambda_2$.
(At present, we have only the upper bound by ACCMOR\cite{ACCMOR}.)

The form factors of the process
 ${\bar B}^0 \rightarrow \pi^+ l {\bar \nu}$ can also be calculated.
There is a $B^*$-pole contribution as in fig.\ref{fig1}.
The form factors are defined as
\begin{equation}
 \langle \pi | J_\mu | {\bar B}^0 \rangle
  = f_+(q^2) (p_B + p_\pi)_\mu + f_-(q^2) (p_B - p_\pi)_\mu,
\end{equation}
 where $q^2 = (p_B - p_\pi)^2$.
We obtain
\begin{equation}
 f_{\pm}(q^2)
 = {1 \over 2} {{f_B} \over {f_\pi}}
   \left[
   1 - {{f_{B^*}} \over {f_B}}
        \left\{
        \lambda \left( 1 + {{v \cdot p_\pi} \over {2M_b}} \right)
        + (\lambda_1 - \lambda_2) {\Lambda \over {M_b}}
        \right\}
       {{2 M_b \left( v \cdot p_\pi \mp m_B \right)}
                              \over {q^2 - m_{B^*}^2}}
   \right],
\end{equation}
 where
\begin{equation}
 v \cdot p_\pi = {{m_B^2 + m_\pi^2 - q^2} \over {2 m_B}}.
\end{equation}
The axial couplings $\lambda_1$ and $\lambda_2$
 are contained in the combination of $\lambda_1-\lambda_2$
 also in this form factor.
This result coincides with the result in ref.\cite{Burdman-et-al}
 in the soft pion limit $p_\pi \rightarrow 0$
 \footnote{The sign and the normalisation of the axial coupling
            $\lambda$ is different from the one
            in ref.\cite{Burdman-et-al}.}.
Our result
 gives an extension of their result
 to $0 \le |p_\pi| \lesssim 4\pi f_\pi$.
(The upper bound $4\pi f_\pi$ is taken
 as the limit of the chiral expansion.)

This form factors are important in extracting $|V_{ub}|$
 from the exclusive decay of $B \rightarrow \pi l \nu$.
The $q^2$ dependence is given,
 which gives valuable information on the axial coupling constant
 by fitting the form factors with the $q^2$ spectrum.

The form factors of the process
 ${\bar B}^0 \rightarrow \rho^+ l {\bar \nu}$ can be calculated.
There is a $B$-pole contribution as in fig.\ref{fig2}.
The form factors are defined as
\begin{eqnarray}
 \langle \rho | J^\mu | {\bar B}^0 \rangle
 &=& f_1(q^2) \epsilon^{\mu\nu\rho\sigma} \epsilon^*_\nu(p_\rho)
              (p_B + p_\rho)_\rho (p_B - p_\rho)_\sigma
\nonumber\\
 &+& f_2(q^2) i \epsilon^{*\mu}(p_\rho)
\nonumber\\
 &+& f_3(q^2) i (\epsilon^*(p_\rho) \cdot p_B) (p_B + p_\rho)^\mu
\nonumber\\
 &+& f_4(q^2) i (\epsilon^*(p_\rho) \cdot p_B) (p_B - p_\rho)^\mu,
\end{eqnarray}
 where $q^2 = (p_B - p_\rho)^2$.
We obtain
\begin{eqnarray}
 f_1(q^2) &=& 0,
\\
 f_2(q^2) &=& f_B g_V - \beta_2 \sqrt{M_b \Lambda} g_V,
\\
 f_3(q^2) &=& {1 \over 2} \beta_1
              {{\sqrt{M_b \Lambda}} \over {M_b^2}} g_V,
\\
 f_4(q^2) &=& 2 f_B g_V \left\{
                         r + {\Lambda \over {M_b}} (r_1 + 6 r_2)
                        \right\} {1 \over {m_{B^\pm}^2 - q^2}}
           +  {1 \over 2} \beta_1
              {{\sqrt{M_b \Lambda}} \over {M_b^2}} g_V.
\end{eqnarray}
As same as in the case of
 ${\bar B}^0 \rightarrow \pi^+ l {\bar \nu}$,
 this result is valid for the low energy $\rho$ meson
 since we are using the chiral expansion.
The form factor $f_4(q^2)$
 is not significant as far as we can neglect the lepton masses.
If we can fix the parameters $\beta_1$, $\beta_2$, $f_B$, and $g_V$,
 the Kobayashi-Maskawa matrix element $V_{ub}$
 can be extracted from this decay mode.

Many precise experiments which are expected in future B-factories
 will be used to fix the parameters in this effective theory.
We have the predictive power once the parameters are fixed.
The prediction is model independent,
 since we only used the information of the symmetry of QCD
 in constructing the effective theory.

In this paper,
 we construct the heavy meson effective Lagrangian
 up to $O(p^2)$ in the chiral expansion
 and $O(1/M_Q^2)$ in the $1/M_Q$ expansion.
We enumerated all the possible terms
 which are allowed by the symmetry.
The reparameterisation invariance in the meson level
 is very important in the construction.
Many terms are dropped out by the invariance.
The light vector mesons are introduced
 by the method of the hidden local symmetry.
The weak current in the effective theory is obtained in the same way.
The decay width of the process $D^{*+} \rightarrow D^0 \pi^+$,
 the form factors of the processes
 ${\bar B}^0 \rightarrow \pi^+ l {\bar \nu}$
 and ${\bar B}^0 \rightarrow \rho^+ l {\bar \nu}$
 are calculated.
The results are easily translated
 to the $D$-meson semileptonic decays.
The meson effective Lagrangian
 can give the simple and physically clear understanding
 on the heavy meson decays.

We are grateful
 to Dr.~M.Tanaka of KEK theory group for helpful advice,
 and Prof.~A.I.Sanda of Nagoya university for encouragement.
T.K. would like to thank Dr.~G.Boyd for fruitful discussion
 on $1/M$ correction and reparameterisation invariance.
The work by him with B.Grinstein, SSCL-preprint-532,
 also deals with the $O(1/M)$ terms in heavy meson Lagrangian.

T.K.'s work
 is supported in part Grant-in Aids for Scientific Research
 from the Ministry of Education,
 Science and Culture (No. 03302015 and 05228104).

\begin{figure}
\caption{The diagrams for
          ${\bar B}^0 \rightarrow \pi^+ l {\bar \nu}$ decay.
         The black circle and box
          represent the strong and weak vertices, respectively.}
\label{fig1}
\end{figure}
\begin{figure}
\caption{The diagrams for
         ${\bar B}^0 \rightarrow \rho^+ l {\bar \nu}$ decay.}
\label{fig2}
\end{figure}


\begin{references}
\bibitem{Isgur-Wise}
 N.Isgur and M.B.Wise, Phys. Lett. B232 (1989) 113; B237 (1990) 527.
\bibitem{Heavy-quark}
 H.Georgi, Phys. Lett. B240 (1990) 447;
 M.J.Dugan, M.Golden, and B.Grinstein, Phys. Lett. B282 (1992) 142.
\bibitem{Heavy-meson}
 H.Georgi, Lectures delivered at TASI,
  Published in Boulder TASI 91, 589 (HUTP-91-A039);
 G.Burdman and J.F.Donoghue, Phys. Lett. B280 (1992) 287;
 M.B.Wise, Phys. Rev. D45 (1992) R2188.
\bibitem{Luke}
 M.E.Luke, Phys. Lett. B252 (1990) 447;
 M.Neubert, Phys. Lett. B306 (1993) 357; SLAC-PUB-6258.
\bibitem{Luke-Manohar}
 M.Luke and A.V.Manohar, Phys. Lett. B286 (1992) 348;
 M.Neubert, Phys. Lett. B306 (1993) 357.
\bibitem{Casalbuoni-et-al-1}
 R.Casalbuoni, A.Deomdrea, N.Di Bartolomeo, R.Gatto,
 F.Feruglio, and G.Nardulli,
 Phys. Lett. B294 (1992) 106;
 R.Casalbuoni, N.Di Bartolomeo, R.Gatto, F.Feruglio, and G.Nardulli,
 Phys. Lett. B299 (1993) 139.
\bibitem{Yan-et-al}
 T-M.Yan, H-Y.Cheng, C-Y.Cheng, G-L.Lin, Y.C.Lin, and H-L.Yu,
 Phys. Rev. D46 (1992) 1148.
\bibitem{Casalbuoni-et-al-2}
 R.Casalbuoni, A.Deomdrea, N.Di Bartolomeo, R.Gatto,
 F.Feruglio, and G.Nardulli,
 Phys. Lett. B292 (1992) 371.
\bibitem{Bando-et-al}
 M.Bando, T.Kugo, S.Uehara, K.Yamawaki, and T.Yanagida,
 Phys. Rev. Lett. 54 (1985) 1215;
 For review, M.Bando, T.Kugo, and K.Yamawaki, Phys. Rep. 164 (1988) 217.
\bibitem{Cheng-et-al}
 H-T.Cheng, C-Y.Cheung, G-L.Lin, Y.C.Lin, T-M. Yan, and H-L. Yu,
 CLNS 93/1192.
\bibitem{CLEO}
 CLEO collabolation, Phys. Rev. Lett. 69 (1992) 2041.
\bibitem{ACCMOR}
 ACCMOR collabolation, Phys. Lett. B278 (1992) 480.
\bibitem{Burdman-et-al}
 M.Neubert, SLAC-PUB-6344;
 G.Burdman, Z.Ligeti, M.Neubert, and Y.Nir, SLAC-PUB-6345.
\end{references}
\end{document}